\documentclass[aapm,graphicx,article,superscriptaddress,amsmath,amssymb]{revtex4-1}

\usepackage[pdftex]{graphicx}
\DeclareGraphicsExtensions{.eps}
\usepackage{epstopdf}
\usepackage{amsthm}
\usepackage{color}
\usepackage{array}
\usepackage{subfigure}
\usepackage{mathtools}
\usepackage[bottom]{footmisc}
\usepackage{algorithm}
\usepackage[noend]{algpseudocode}
\usepackage{tabularx}
\usepackage{booktabs} 
\usepackage[mathlines]{lineno}

\begin{document}

\preprint{AAPM/123-QED}

\title[Adaptation Cost and Tractability in Robust ART]{On Adaptation Cost and Tractability in Robust Adaptive Radiation Therapy Optimization} 
\thanks{There is no COI or financial disclosure to report.}

\author{Michelle B\"ock}
\email{miboeck@kth.se}
\affiliation{KTH Royal Institute of Technology, Stockholm, Sweden}
\affiliation{RaySearch Laboratories AB, Stockholm, Sweden}
\date{\today}

\begin{abstract}

\textbf{Purpose:} 
In this paper, a framework for online robust adaptive radiation therapy~(ART) is discussed and evaluated.
The purpose of the presented approach to ART is to: (i) handle interfractional geometric variations following a probability distribution different from the \textit{a priori} hypothesis, (ii) address adaptation cost and (iii) address computational tractability.

\textbf{Methods:}
A novel framework for online robust ART using the concept of Bayesian inference and scenario-reduction is introduced and evaluated in a series of treatment on a one-dimensional phantom geometry.
The initial robust plan is generated from a robust optimization problem based on either expected-value- or worst-case-optimization approach using the \textit{a priori} hypothesis of the probability distribution governing the interfractional geometric variations.
Throughout the course of every treatment, the simulated interfractional variations are evaluated in terms of their likelihood with respect to the \textit{a priori} hypothesis of their distribution and violation of user-specified tolerance limits by the accumulated dose.
If an adaptation is considered, the \textit{a posteriori} distribution is computed from the actual variations using Bayesian inference. Then, the adapted plan is optimized to better suit the actual interfractional variations of the individual case. 
This adapted plan is used until the next adaptation is triggered. 
To address adaptation cost, the proposed framework provides an option for increased adaptation frequency.
Computational tractability in robust planning and ART is addressed by approximation algorithms to reduce the size of the optimization problem. 

\textbf{Results:} 
According to the simulations, the proposed framework may improve target coverage compared to the corresponding non-adaptive robust approach. In particular, combining the worst-case-optimization approach with Bayesian inference may perform best in terms of improving CTV coverage and organ-at-risk~(OAR) protection. 
Concerning adaptation cost, the results indicate that mathematical methods like Bayesian inference may have a greater impact on improving individual treatment quality than increased adaptation frequency.
In addition, the simulations suggest that the concept of scenario-reduction may be useful to address computational tractability in ART and robust planning in general. 

\textbf{Conclusion:}
The simulations indicate that the adapted plans may improve target coverage and OAR protection at manageable adaptation and computational cost within the novel framework. 
In particular, adaptive strategies using Bayesian inference appear to perform best among all strategies.
This proof-of-concept study provides insights into the mathematical aspects of robustness, tractability and ART, which are a useful guide for further development of frameworks for online robust ART.


\end{abstract}

\maketitle 

\section{Introduction}
In adaptive radiation therapy~(ART), the treatment plan is adapted in order to account for interfractional geometric variations in the individual patient.
These changes and their impact on the accumulated dose can be monitored in the course of image-guided radiation therapy~(IGRT). 
Traditionally, interfractional geometric variations are handled by adding safety margins around the organs-at-risk~(OARs) and clinical target volume~(CTV).  
The CTV is expanded to a larger planning target volume~(PTV) by using population based margin-recipes such as presented by Stroom \textit{\textit{et al.}}~\cite{stroom1999} and van Herk \textit{et al.}~\cite{vanHerk2000,vanHerk2004}.
This PTV is then irradiated with the prescribed curative dose.
However, the PTV-concept has limitations which lead to research on robust treatment planning using the concepts of stochastic- and minimax-optimization~\cite{unkelbach2018}. 
These methods directly take into account the presence of interfractional variations into the optimization process.
In particular, treatment modalities which are sensitive to interfractional variations, such as intensity modulated radiotherapy~(IMRT) and intensity modulated proton therapy~(IMPT) seem to benefit from robust planning~\cite{unkelbach2018}.
Typically, these variations are modeled as discrete scenarios with corresponding probabilities according to an \textit{a priori} estimate of their distribution~\cite{unkelbach2018}. 
The resulting treatment quality relies on the \textit{a priori} hypothesis on distribution and the order of magnitude of the interfractional variations. 

However, in the event of interfractional variations that are distributed differently than anticipated in the \textit{a priori} hypothesis, applying the same plan throughout the course of treatment may impair treatment quality~\cite{yan1997}. Thus, ART may improve treatment quality~\cite{bock2017,hoffmann2017,limreinders2017}. 
In general, the planning process is time intensive and requires a cross-disciplinary team of radiation oncologists, dosimetrists, therapists and medical physicists. 
If a plan has to be adapted, the contours of the tumor(s) and OARs may be adjusted, the plan reoptimized, approved by the radiation oncologist and reexamined for patient-specific quality assurance~(QA)~\cite{limreinders2017}. 
Typically, ART is categorized into online and offline ART.
In online ART, the plan is adapted with the patient lying in treatment position, which limits the time available for replanning and reexamining the adapted plan.
In offline ART, the plan is adapted and examined in between fractions. 
In general, ART is considered expensive in terms of time and resources such as workforce and computation time. 
Therefore, ART is not commonly used in conventional clinical routine.
Despite obstacles in workflow implementation, ART has been the subject of various theoretical and practical studies on clinical patient data.
Theoretical studies focus on the problem formulation and optimization approach. 
According to our previous studies~\cite{bock2017,bock2018}, the concept of combining robust optimization approaches with adaptive replanning can handle interfractional geometric variations.
In particular, this combination can outperform the conventional non-adaptive treatment approach in terms of CTV coverage and OAR protection. 
Moreover, the connection between more or less conservative approaches to robust planning and optimization variables which are either time-and/or-scenario dependent or static is investigated. 
On the contrary, the majority of theoretical studies use radiobiological models to adapt and personalize the treatment schedule and fraction size~\cite{arcangeli2002,kim2012,kim2015,saberian2016}.   
However, radiobiological models are considered uncertain and therefore hardly used in clinical practice~\cite{bentzen2010,allenli2012}.
Practical studies are carried out on a sample of in-house treated patients in order to study the potential of ART through reoptimization~\cite{schwartz2013,limreinders2017} or a plan-of-the-day approach~\cite{hafeez2017,limreinders2017}, and how to manage the increasing workload. 
These studies emphasize the importance of identifying those patients who may benefit the most from ART and determining the appropriate adaptation frequency.
Moreover, the deficiencies of the current conventional approach to planning, time limitations and computational tractability are discussed.
However, these studies do not take a systematic approach to ART planning. 

In our previous work~\cite{bock2017,bock2018}, robust optimization approaches were combined with adaptive reoptimization to establish mathematical models of frameworks for robust ART.  
These models were evaluated in simulation and mathematical studies in the context of offline robust ART.
In this work, the findings of the previous studies are used to extend our research in mathematical modeling to design a framework for online robust ART.
Here, a framework for online robust ART is presented and evaluated on a one-dimensional phantom geometry for a series of simulated treatments.
The aim of this proof-of-concept study is to identify the relevant mathematical properties of such a framework which is supposed to handle the following issues. 
First, inaccuracies in the \textit{a priori} estimate of the probability distribution of the interfractional geometric variations are handled by employing Bayesian inference to compute the \textit{a posteriori} distribution of the actually occuring variations for each individual case.
Second, the issue of computational tractability is addressed by using approximation algorithms to reduce the size of the robust optimization problem. 
Third, the issues of adaptation costs and the appropriate adaptation frequency are addressed by the presented adaptive strategies and mathematical adaptation triggers.
By evaluating the proposed framework on a one-dimensional phantom geometry, features in the resulting dose distributions can be associated with the corresponding optimization parameter. 
This allows us to gain insights into the relevant mathematical properties for an effective framework for robust ART~\cite{lof1998,chan2006,fredriksson2014,bock2017,bock2018}.
Such insights will be valuable for further development of this framework.

\section{Methods}\label{sec:Methods}
In this section, we present the mathematical foundation of the proposed framework for online robust ART.
First, the novel concept is presented and motivated. 
Second, the concept of Bayesian inference is introduced and its use in the proposed framework is described. 
Third, the use of an algorithm for scenario-reduction and tree-generation from the literature to address the issue of computational tractability is motivated.
Fourth, the three adaptive strategies.
Fifth, the two robust optimization approaches employed in the framework are discussed.

\subsection{The Proposed Framework for Online Robust ART} \label{sec:framework}
\begin{figure}
\centering
\subfigure[Schematic planning process and represenation of the conventional non-adaptive robust treatment strategy.]{\includegraphics[scale=0.55]{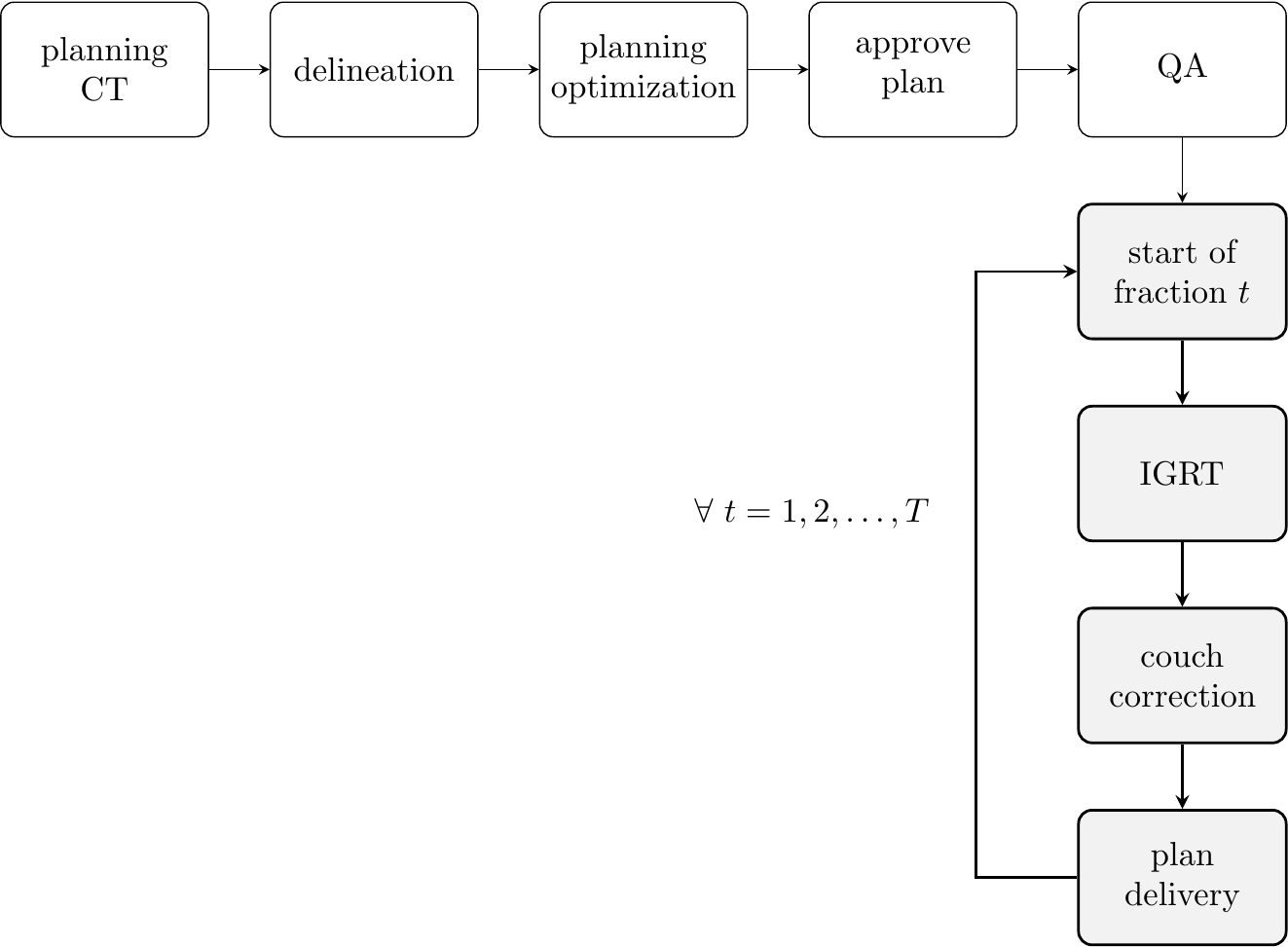} \label{fig:NA_chart}}\quad 
\subfigure[Representation of our robust adaptive framework in an online setting.]{\includegraphics[scale=0.55]{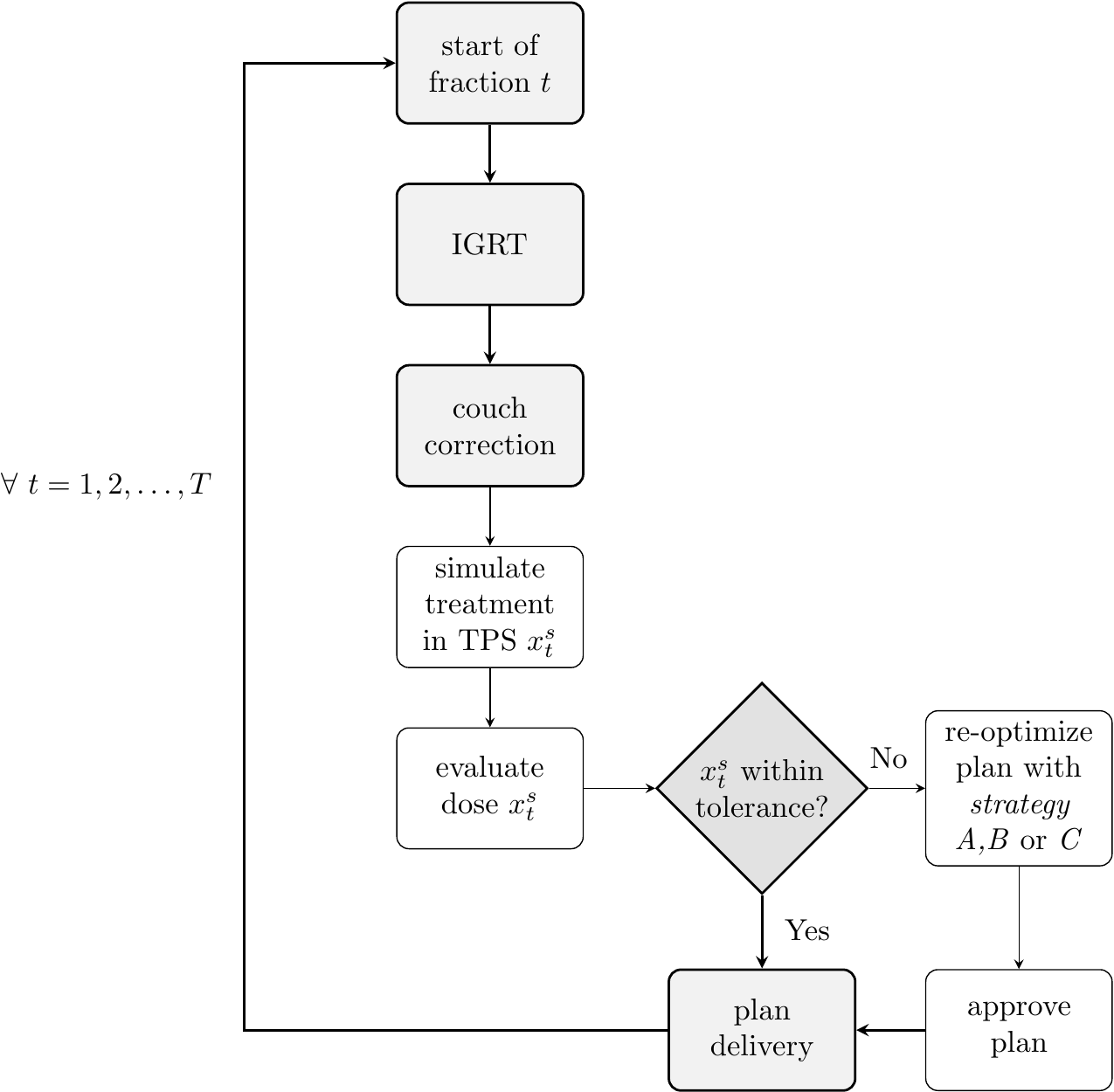} \label{fig:ART_Chart}}
\caption{Comparison between the non-adaptive robust treatment strategy and our proposed robust adaptive framework to support the decision-making process. The white boxes represent actions which are performed in the treatment planning system, while the grey boxes represent actions performed while the patient is lying on the treatment couch.}
\label{fig:NAvsART}
\end{figure}

In this online framework, see Figure~\ref{fig:ART_Chart}, treatment starts with an initial robust plan. 
This initial robust plan is the optimal solution of the robust optimization problem which accounts for interfractional variations according to a population-based \textit{a priori} hypothesis of their distribution.
During treatment, the actual interfractional variations are evaluated in terms of their likelihood with respect to the \textit{a priori} hypothesis of their distribution and their impact on the accumulated dose with respect to user-specified criteria.
Thus, both mathematical and clinical criteria may trigger adaptive replanning.
It is assumed that the actually occuring interfractional variations and their impact on the accumulated dose is measured  before plan delivery as part of the daily IGRT-routine.
Here, systematic errors are assumed to be corrected during the IGRT-routine.
If an adaptation is triggered, parameters in the robust optimization problem are updated in response to the measured interfractional variations using Bayesian inference. 
This modified robust optimization problem is then solved in order to generate the adapted plan. 
Typically, any plan has to be approved by an oncologist before delivery.
Here, this process is handled by evaluating the improvement of objective function value. 
Then, the adapted plan is delivered in the current and the subsequent fractions until the next adaptation will be triggered.

\subsection{Bayesian inference}\label{sec:bayes}
In this framework, the concept of Bayesian inference is applied in the following manner.
The interfractional geometric variations are modeled to be independent and identically distributed (i.i.d.) according to a normal distribution~$\mathcal{N}(0,\sigma^2_{\text{prior}})$, where the mean is assumed to be zero. 
The standard deviation represents the amplitude of the interfractional variations and is considered to be case-specific. 
Due to the lack of case-specific data, a population-based estimate~$\sigma_{\text{prior}}$ is used.
During the course of treatment case-specific data become available, denoted by~$\bar{\xi}$, through the daily monitoring of the interfractional variations.
If the plan is adapted at~$t_P$, the following measurements~$\bar{\xi} = \{\xi_{1},\dots,\xi_{t_P} \}$ of the individual case are available.
In case an adaptation has taken place at a previous time instant~$t^p_P$, the sample will be denoted by~$\bar{\xi} = \{ \xi_{t^p_P},\dots,\xi_{t_P} \}$. 
Given our measurements~$\bar{\xi}$, the \textit{a posteriori} distribution~$\pi(\sigma^2|\bar{\xi})$ of the case-specific variance~$\sigma^2$ is computed from
\begin{equation}\label{eq:posterior}
\pi(\sigma^2|\bar{\xi}) = \frac{\pi(\bar{\xi}|\sigma^2)\pi(\sigma^2)}{\pi(\bar{\xi})}.
\end{equation}
The \textit{a posteriori} distribution~$\pi(\sigma^2|\bar{\xi})$ 
is a conditional distribution of the variance given the data~$\bar{\xi}$, where~$\pi(\bar{\xi}|\sigma^2)$ denotes the likelihood of the measurements given the \textit{a priori} distribution~$\pi(\sigma^2)$.
Moreover,~$\pi(\bar{\xi})$ denotes the marginal distribution  which gives the probability  that the data~$\bar{\xi}$ are measured.
Here, the a posteriori distribution is used to compute the point estimate~$\mathbb{E}[\sigma^2|\bar{\xi}]$ of the case-specific variance. 
We follow Casella and Berger~{\cite[Ch.\ 7,p.\ 359]{casellaberger2001}} by choosing the inverted gamma distribution~$\text{IG}(\alpha,\beta)$ as the \textit{a priori} distribution of~$\sigma^2$.
Taking into account our measurements with sample size~$
k=|\bar{\xi}|$, we get the following posterior distribution of $\sigma^2$,~$\text{IG} \left(  \alpha + \frac{k-1}{2},\left(  \frac{S_k^2(k-1)}{2}+\frac{1}{\beta}\right)^{-1}  \right)$ where~$S_k$ denotes the unbiased variance~$S_k^2 = \frac{1}{k-1}\sum_{i=1}^k(\xi_i -\mu)^2$, and~$\alpha = 1$ and~$\beta = 1$. 
Then, the point estimate~$\mathbb{E}[\sigma^2|\bar{\xi}]$ of the case-specific variance is given by
\begin{equation}\label{eq:post_var}
\sigma^2_{\text{post}}=\mathbb{E}[\sigma^2|\bar{\xi}] = \frac{S_k^2 + \frac{2}{\beta(k-1)}}{1 + \frac{2(\alpha-1)}{k-1}},
\end{equation}
which is used during adaptive  reoptimization in the proposed framework.

\subsection{Tractability and Scenario Reduction}\label{sec:scenred}
In robust planning, interfractional variations are typically modeled as discrete scenarios~\cite{unkelbach2018}.
The number of all possible scenario realizations grows exponentially with the number of fractions.
As a consequence, optimization problems including all scenarios may become expensive to solve or even intractable. 
In this framework, the issue of tractability is addressed by applying the concept of scenario-reduction and scenario-tree generation to robust ART. 
In general, scenario reduction algorithms approximate the large original problem by a subset of scenarios of a predefined cardinality or accuracy~\cite{vazsonyi2006,dupacova2003,growe2003}. 
Such methods are commonly used in decision making models in portfolio and risk managment for electrical power utilities that deal with uncertainty~\cite{dupacova2003,growe2003}.
To the best of our knowledge, this is the first application of such methods to robust planning and ART.
Here, we choose to apply the scenario-reduction- and tree-generation algorithm proposed by~Gr\"{o}we-Kuska \textit{\textit{et al.}}~\cite{growe2003}, which is well-suited to handle the issue of intractability in this paper.
The concept of the algorithm used in this framework is presented in pseudo-code in Algorithm~1 and~2 and described as follows.
This algorithm stepwise reduces the number of nodes in a fan of individual scenarios, as similar scenarios are bundled together and the tree structure is modified.
To determine whether a scenario is discarded or preserved, the Kantorovich distance~{\cite[p.\ 427-430]{memoli2011}} is used to trade off scenario probabilities and distances of scenario values. 
In this framework, the decision maker may specify the number of preserved scenarios at fraction~$T$. 
The scenarios to be discarded are identified by recursively applying a \textit{maximum reduction strategy}~(MRS)~{\cite[eq.(2)]{growe2003}} from~$T,T-1,\dots,1$. 
Given the accuracy~$\epsilon$ with which the probability distribution of the reduced set approximates the probability distribution of the original set, the index set~$J$ of maximal cardinality is determined by
\begin{equation}\label{eq:MRS}
\sum_{i \in J} p_i \ \underset{ j \notin J }{\mathrm{min}} \ c_T \left( \omega^i,\omega^j \right) \leq \epsilon.
\end{equation}
The distance between scenarios,~$\omega^i$ and~$\omega^j$, over the total time horizon is denoted by $c_T(\omega^i,\omega^j) \coloneqq \sum_{\tau = 1}^T |\omega^i_{\tau}-\omega^j_{\tau}|$.
Thus,~$J$ contains the indices of the discarded scenarios. 
The preserved scenarios~$\omega^j,j\notin J$ are assigned new probabilities~$q_j,q_j\geq 0,\sum_j q_j = 1$, according to the \textit{optimal redistribution rule}~{\cite{growe2003}}
\begin{equation}\label{eq:RedistRule}
q_i \coloneqq p_j + \sum_{i \in J(j)} p_i, \text{where} \ J(j) \coloneqq \{i\in J:j=j(i) \}, j(i) \in \mathrm{arg} \ \underset{j \notin J}{\mathrm{min}}c_T \left( \omega^i,\omega^j \right), \ \forall \in J.
\end{equation}
According to the rule given by~(\ref{eq:RedistRule}), the new probability of a preserved scenario sequence is equal to the sum of its former probability~$p_j$ and of all probabilities of the deleted scenarios,~$\sum_{i \in J(j)} p_i$, that are closest to it with respect to~$c_T \left( \omega^i,\omega^j \right)$.

\begin{figure*}
\center
\includegraphics[scale=0.54]{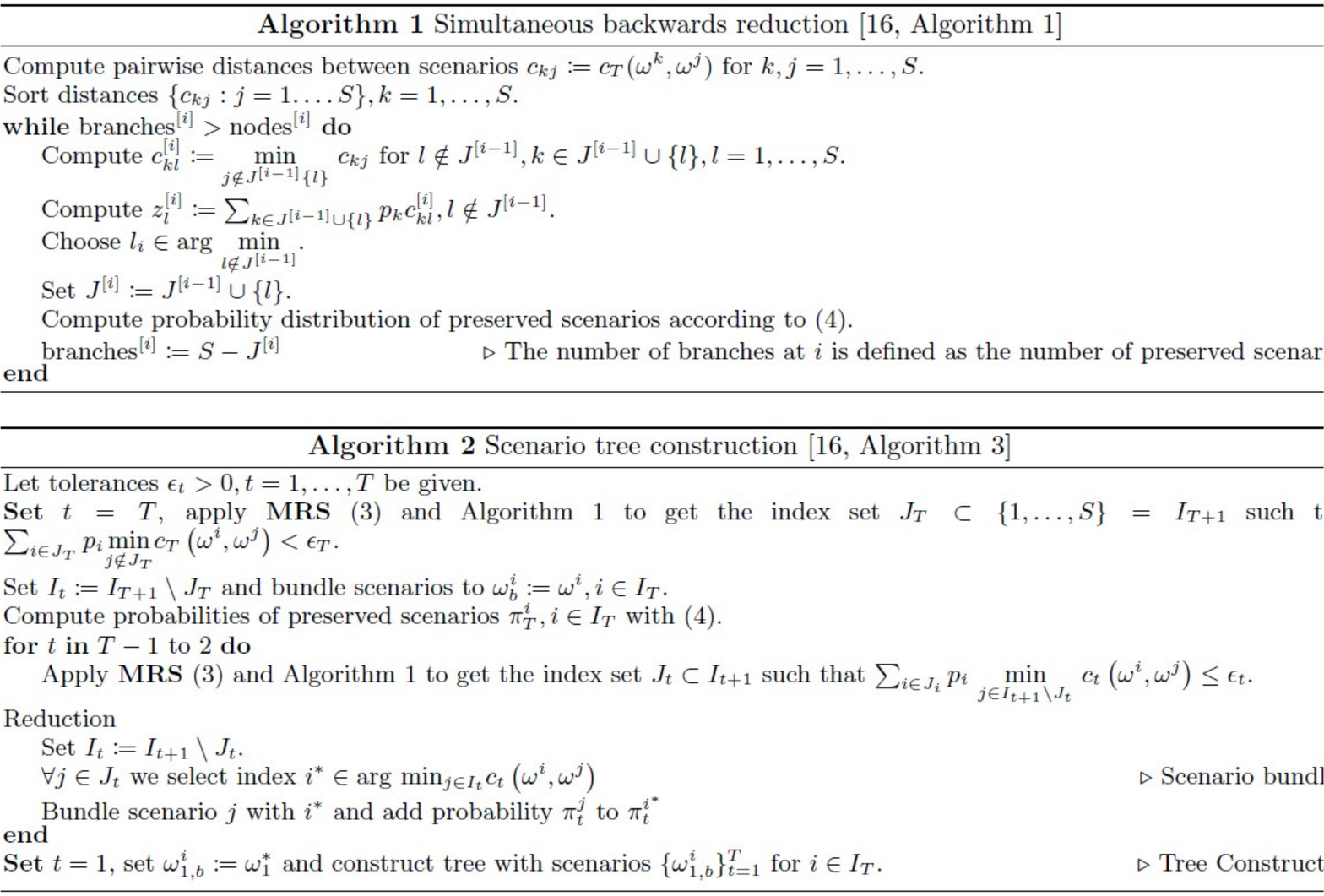}
\end{figure*}

\subsection{Adaptive Strategies} \label{sec:strategies}
The adaptive strategies in this proposed framework are designed to address the issues of adaptation costs and appropriate adaptation frequency. 
To evaluate the interfractional variations and their impact on the accumulated dose in a straightforward manner, mathematical and user-specified adaptation triggers are introduced. 
For the purpose of performing simulations in this study, dose-volume-points are chosen as user-specified adaptation triggers.

The user-specified criteria used in the simulatios follow recommendations for a prostate case in the ICRU83 report~\cite{ICRU83}.
In terms of OAR protection,~25\% of the bladder and~35\% of the rectal wall should not receive more than~32\% and~45\% of the prescribed dose, respectively.
Here, the~$D_{25}$ and~$D_{35}$ criteria are used for the right and left OAR, respectively.
Concerning the mathematical adaptation triggers, the actual interfractional variations are evaluated in terms of their likelihood with respect to the current estimate of their probability distribution.
If more than half of the actual variations have a likelihood of less than~60\%, adaptation will be triggered followed by computing the
\textit{a posteriori} estimate of case-specific standard deviation with Bayesian inference.
Here, the threshold of~60\% is chosen to account for deviations larger than one standard deviation for the majority of the simulated treatments. 
However, this threshold can be chosen by the decision-maker.
Then, the potential improvement of the adapted plan is assessed by comparing the objective function value evaluated at the adapted and previous plan for approval. 
For this purpose, the objective function is parameterized with the current estimate of case-specific standard deviation.
In this framework, three adaptive strategies are introduced and evaluated which are referred to as \textit{strategy A}, \textit{B} and \textit{C}. All three strategies share the approach to take into account the already delivered dose and remaining number of fraction.
\begin{itemize}
\item In \textit{strategy A}, the adapted plan is optimized with respect to the remaining dose to be delivered and the remaining numbers of fractions.
Treatment plan adaptation can be performed whenever adaptation has been triggered.

\item In \textit{strategy B}, Bayesian inference is used to compute the case-specific standard deviation.
Then the adapted plan is optimized with respect to the case-specific standard deviation, remaining dose and time-horizon.
However, the plan is adapted only if a sufficient amount of data~$\bar{\xi}$ is available, leading to a delay in between adaptations. 
Here, the delay is set to five fractions corresponding to one week of a six week long treatment.
It is the goal to evaluate potential improvements by adapting plans using Bayesian inference. 
For that purpose, \textit{strategy A} is a lower benchmark. 

\item In \textit{strategy C}, Bayesian inference is used to compute the case-specific standard deviation.
In contrast to \textit{strategy B}, an adapted plan can be optimized whenever an adaptation is triggered. 
In other words, if the previous adaptation lies less than five fractions behind, the plan is adapted with respect to the current estimate of the case-specific standard deviation like in \textit{strategy A}. 
Otherwise, if the previous adaptation lies more than five fractions behind, the case-specific standard deviation is updated by using Bayesian inference followed by optimizing the adapted plan like in \textit{strategy B}. 
It is the goal to evaluate the extent of potential improvements by combining the option for frequent adaptations with Bayesian inference.
\end{itemize}
In order to evaluate the potential improvements of the proposed framework for robust ART, the non-adaptive robust approach is used as a lower benchmark to evaluate the resulting dose distribution at the end of the treatment.
By combining mathematical, user-specified and the three adaptive strategies, the issues of adaptation costs and adaptation frequency may be handled in a systematic and less biased fashion.

\subsection{Robust Planning Models}
Here, robust optimization approaches are used to optimize the initial and adapted robust plans.
In our previous studies, expected-value-, worst-case-and conditional-value-at-risk~(CVaR) optimization was used to evaluate robust approaches with varying grades of conservativeness. 
The results with the clearest trends in the previous studies were given by expected-value- and worst-case-optimization.
Therefore, the expected-value- and worst-case-optimi\-zation approach are used in this framework.  
While expected-value-optimization is considered the least conservative, worst-case-optimization is the most conservative approach.
In general, the proposed framework can be combined with any other robust optimization approach.
The robust plans~$u$ are optimized to handle interfractional variations such that the delivered final dose~$x_T$ resembles the prescribed dose distribution~$d_T$ as well as possible.
This goal is expressed through the function~$g(x_T,d_T)$ chosen by the decision-maker. The details to the particular function~$g(x_T,d_T)$ used in this proof-of-concept study is discussed later. 
Here, the expected-value-optimization approach is given by
\begin{equation}\label{eq:exp}
\underset{u\geq 0}{\mathrm{min}} \ \mathbb{E}_{\sigma}\left[ g(x_T(u,\bar{\omega}_1^T),d_T) \right],
\end{equation}
while the worst-case-optimization approach is formulated to
\begin{equation}\label{eq:WC}
\underset{u\geq 0}{\mathrm{min}} \ \underset{ {\bar{\omega}_1}^T \in \{\Omega_{\sigma} \times \cdots \times \Omega_{\sigma}\}_T }{\text{max}} g(x_T(u,\bar{\omega}_1^T),d_T),
\end{equation}
where the subscript~$\sigma$ represents the parametrization of the objective function with respect to the current estimate of the case-specific standard deviation; and~$ \bar{\omega}_1^T$ denotes the possible scenario realizations from fraction one to T. All possible scenario realizations are summarized in the set~$\{\Omega_{\sigma} \times \cdots \times \Omega_{\sigma}\}_T$ of the cardinality~$|\Omega|^T$. The set~$\Omega$ contains the scenarios~$\omega$ which may occur at every fraction.

\section{Computational Study} \label{sec:experiment}
\subsection{Optimization Formulation and Phantom Geometry}\label{sec:expstduy}
In this proof-of-concept study, the dose criterion chosen as the objective function~$g(x_T,d_T)$ penalizes the deviation from a uniform dose of one in the CTV and a dose of zero elsewhere according to
\begin{equation}\label{eq:quadpen}
\sum_{r\in\mathcal{R}} w_r \sum_{n\in\mathcal{N}_r}  v_{n,r}(x(u,\bar{\omega}_1^T)_{T,n}-d_{T,n})^2,
\end{equation}
where~$n$ denotes the voxel index contained in the index set~$\mathcal{N}_r$ for every region of interest~(ROI)~$r \in \mathcal{R}$, denoting the set of all ROIs. 
The conventionally used important weights and the relative volumes are denoted by~$w_r$ and~$v_{n,r}$, respectively. The relative volumes in every ROI satisfy~$\sum_{n\in \mathcal{N}_r} v_{n,r} = 1$. 
To emphasize the mathematical properties of this framework, a convex objective function as given by~\eqref{eq:quadpen} is chosen to exploit its smoothness, convexity and the resulting globally optimal solutions of~(\ref{eq:exp}) and~(\ref{eq:WC}).  
Thus, the resulting initial and adapted robust plans can be interpreted in a straightforward manner. 
In general, any type of objective function can be combined with the proposed framework. 
However, the characteristics such as convexity are useful to study novel frameworks from a mathematical perspective, as done by~Unkelbach and Oelfke~\cite{unkelbach2004} and Fredriksson~\cite{fredriksson2012} among others~\cite{unkelbach2009,bock2017,fredriksson2016,bokrantz2017}. 
In addition, the expected value of the quadratic penalty~(\ref{eq:exp}) has advantageous in terms of computational tractability. 
The objective function~$ \mathbb{E} \left[ \sum_{r\in\mathcal{R}} w_r \sum_{n\in\mathcal{N}_r}  v_{n,r}(x(u,\bar{\omega}_1^T)_{T,n}-d_{T,n})^2 \right]$ can be reformulated by exploiting its structure and the assumption of i.i.d. interfractional variations, which gives
\begin{equation}\label{eq:quadpen2}
\sum_{r\in\mathcal{R}} w_r \sum_{n\in\mathcal{N}_r}  v_{n,r} \left[
\frac{1}{T} \mathrm{Var}[x(u,\omega)_n] + \left( T\mathbb{E}[x(u,\omega)_n] - d_{T,n} \right)^2
\right].
\end{equation}  
Then, the optimal plan is computed from~\eqref{eq:quadpen2} independent of the number of fractions, as demonstrated in our previous work~\cite{bock2018}.  
However, worst case optimization problems such as~(\ref{eq:WC}) are intractable due to an exponential growth in scenarios~\cite{fredriksson2012}. 
Instead, a heuristic is used to find an approximative solution.
In our framework, we apply the scenario reduction and tree generation algorithm, presented in section~\ref{sec:scenred}, to create a tractable problem from which an approximate solution can be obtained.
To allow for spatial and temporal variations in the plans, constraints are used in the optimization problem, such that the daily fraction may vary in the interval of~$(1-\gamma)\frac{d_T}{T} \leq Bu_t \leq (1+\gamma)\frac{d_T}{T}$, where~$\gamma \geq 0$ specifies the user-defined lower and upper bound on the daily dose. 

\begin{figure}
\begin{center}
\includegraphics[scale=0.8]{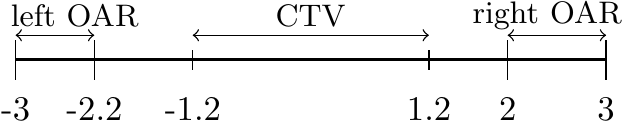}
\caption{One-dimensional patient phantom. The dimensions are given in cm.}\label{fig:Geometry}
\end{center}
\end{figure}

As in our previous study~\cite{bock2018} the proposed framework is evaluated on a one-dimensional phantom geometry, see Figure~\ref{fig:Geometry}. 
This idealized geometry contains one CTV, located between -1.2 and 1.2 cm, and two OARs which are in the intervals of~$[-2,-2.2]$~cm and~$[2,3]$~cm. The OARs are placed asymmetrically around the CTV.
The geometry in question schematically represents both a slice of a two-dimensional phantom and an intersection of a sagittal and transversal cut of a three-dimensional patient-geometry. 
Here, the geometry is discretized into~$N = 60$ voxels.
The interfractional variations are modeled as translational whole-body shifts which can be interpreted as movements in the anterior or posterior direction.
These shifts are included in the optimization process as i.i.d. discrete scenarios which are summarized in the set~$\Omega = \{ \omega_1,\omega_2,\dots,\omega_U \}$, where~$U$ denotes the cardinality of~$\Omega$. 
These scenarios are derived from a discretized normal distribution~$\mathcal{N}(0,\sigma^2_{\text{prior}})$. 
In this study,~$\sigma_{\text{prior}}$ is set to~$0.5$~cm which agrees with the range of movements in the anterior-posterior direction reported in the literature for the prostate~\cite{balter1995,wu2001,langen2001}, cervix CTV~\cite{taylor2008,tyagi2011} and head and neck tumors~\cite{bruijnen2019}.
In this framework, the accumulated dose~$x_t$ at fraction~$t$ is computed according to
\begin{equation}\label{eq:dose}
x_t = x_{t-1} + B S(\omega_t)u \quad \text{for} \ t=1,\dots,T. 
\end{equation} 
The translational shifts are modeled by the socalled shift-operator matrix~$S(\omega_t)$ which corresponds to a translational shift of the plan. 
Since rigid translations are considered only, the shift of a plan is equivalent to a translational shift of the patient in the opposite direction.
Dose deposition in the geometry is modeled by the dose-deposition matrix~$B$.
Before the start of the treatment the patient is not supposed to receive dose and therefore,~$x_0$ is set to zero.
By subjecting this idealized geometry model to rigid translations with respect to the isocenter, dose computation, tracking of the accumulated dose and reoptimization of the adapted plan can be performed in a straightforward manner and computational effort is reduced. 
As a consequence, features in the accumulated dose can be directly associated with parameters in the proposed mathematical model.
This idealized setting has been useful for focusing on the mathematical perspective of optimization for ART in our previous papers~\cite{bock2017,bock2018}.
In the literature, examples of the use of one-dimensional phantom geometries to evaluate the mathematical properties of a novel approach are given by Chan \textit{et al.}~\cite{chan2006} among others~\cite{lof1998,fredriksson2014}.

\subsection{Dose Calculation, Notation and Numerical Optimization}
Here, the radiation plans are delivered by a perpendicular oriented field and the absorbed dose in each voxel is modeled by Gaussian point-spread functions at a spacing of~1~mm with a standard deviation of~3~mm, which are represented by the matrix~$B$ in~(\ref{eq:dose}). 
To reduce computational effort, the number and positions of bixels~$M$ are considered identical with those of the voxels~$N$. 
Thus, the dimensions of the accumulated dose, radiation plans, shift-matrix and dose dose-deposition matrix can be simplified to~$x_t \in \mathbb{R}^N$,~$u \in \mathbb{R}^N$,~$S(\omega)\in \mathbb{R}^{N \times N}$ and ~$B\in \mathbb{R}^{N \times N}$, respectively.
The experimental study is performed using MATLAB~9.4 and the IBM optimization solver CPLEX in the studio version~12.8.
The optimization weights in~(\ref{eq:quadpen}) for the CTV, both OARs and external are set to 100,~10 and~1, respectively.

\subsection{Sensitivity Analysis, Scenario Reduction and Framework Evaluation}

First, a sensitivity analysis is conducted with the goal to evaluate changes in the objective function value of~(\ref{eq:exp}) and~(\ref{eq:WC}) as a function of the standard deviation.  
Here, the objective function value is a measure of treatment quality.
In other words, this analysis is conducted to evaluate how treatment quality is affected, if the distribution of the actual interfractional variations differ from the \textit{a priori} hypothesis.
For this purpose, the unconstrained robust optimization problems~(\ref{eq:exp}) and~(\ref{eq:WC}) are parameterized with respect to standard deviations in the interval of~$[0.0001,1]$. 
Then, an optimal solution~$u^*(\sigma)$ is computed for every value in the interval to obtain the corresponding objective function values~$f_{\sigma}(u^*(\sigma))$. 
These objective function values~$f_{\sigma}(u^*(\sigma))$ represent the ground truth.
The deviation from the ground truth is evaluated for plans generated for four \textit{a priori} estimates~$\sigma_{\text{prior}} = \{ 0.1,0.3,0.5,0.7 \}$.

Second, the suitability of the scenario-reduction and tree-generation algorithm for the proposed framework is investigated.
To address computational tractability, the algorithm is combined with worst-case-optimization. 
The original scenario tree contains~$100.000$ scenarios randomly generated from the \textit{a priori} hypothesis~$\mathcal{N}(0,\sigma^2_{\text{prior}})$. 
The scenario-reduction algorithm is applied to reduce the original tree to smaller trees with the following number of scenarios~$\mathcal{S}_T = \{10.000,5000,1000,500,100\}$.
In this evaluation, the plans obtained from the original tree is compared to the plans obtained from the reduced trees.

Third, the proposed framework is evaluated in its ability to handle inaccurately estimated interfractional variations. 
For this purpose,~200 simulated treatments, each consisting of~30 fractions, are performed on the one-dimensional geometry. 
The population of simulated treatment is generated by discretizing the normal distribution with a standard deviation modeled as a random variable. 
The standard deviation for each treatment is taken from the uniform distribution~$\mathrm{U}[0.2,0.7]$~cm.
Such a wide range of values has been documented for interfractional variations in the literature~\cite{balter1995,wu2001,langen2001,taylor2008,tyagi2011,bruijnen2019}.
A cut-off is applied such that~$95\%$ most probable scenarios in the joint distribution are included and the distribution is renormalized. 
To ensure a fair evaluation of the proposed framework with a variety of robust adaptive strategies and comparison with the corresponding non-adaptive robust strategies, this population of 200 treatments is used.

The statistical accuracy of the computational study is evaluated by using the bootstrap resampling method. 
Bootstrap resampling is a statistical method which is used in studies conducted on a limited number of patient cases~\cite{peters1993,damico2003}. 
This method builds on the basic idea to randomly draw samples with replacement from an original dataset. 
These drawn samples are supposed to be of the same size as the original dataset. 
By repeating this process~$k$ times,~$k$ bootstrap data sets will be generated.
In this study, the bootstrapping is applied to investigate the spread of mean and absolute value of the largest variations in the original data set.
This resampling procedure is carried out~$1000$ times to generate~$1000$ bootstrap data sets.
From these datasets, the probability density estimate is computed for the mean and standard deviation of each dataset.
This evaluation show a narrow spread in the distribution of our bootstrap sample mean and standard deviation, which indicates that the population of 200 treatments is representative to draw valid conclusions from this analysis of the proposed framework.

\section{Results}\label{sec:results}
\subsection{Sensitivity Analysis}
\begin{figure}
\center
\subfigure[Optimal plans~$u^*(\sigma)$ generated by solving the unconstrained problem~(\ref{eq:exp}) for a variety of~$\sigma$.]{\includegraphics[scale=0.35]{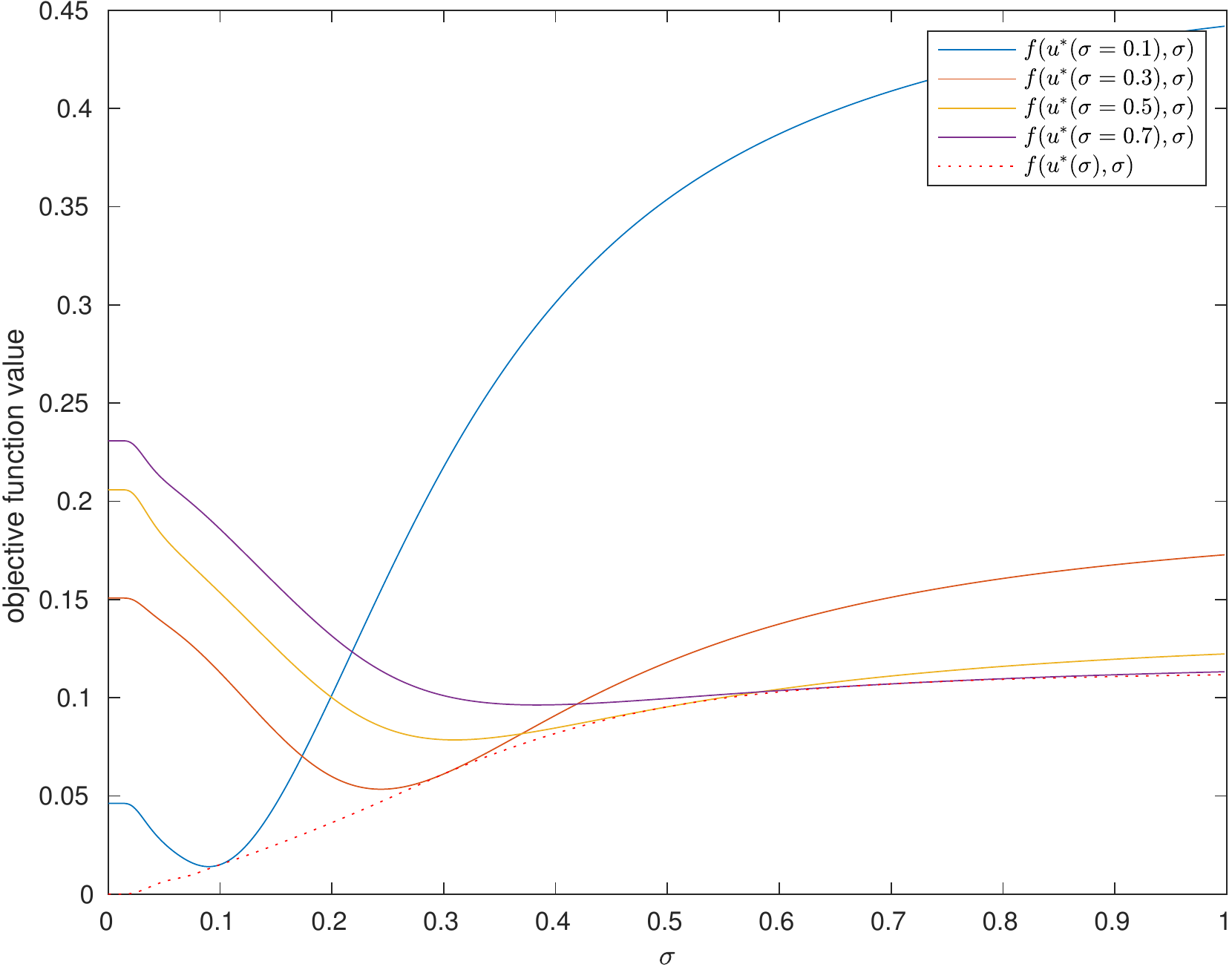}{\label{fig:exp_sens}}
} \quad
\subfigure[Optimal plans~$u^*(\sigma)$ generated by solving the unconstrained problem~(\ref{eq:WC}) for a variety of~$\sigma$.]{\includegraphics[scale=0.35]{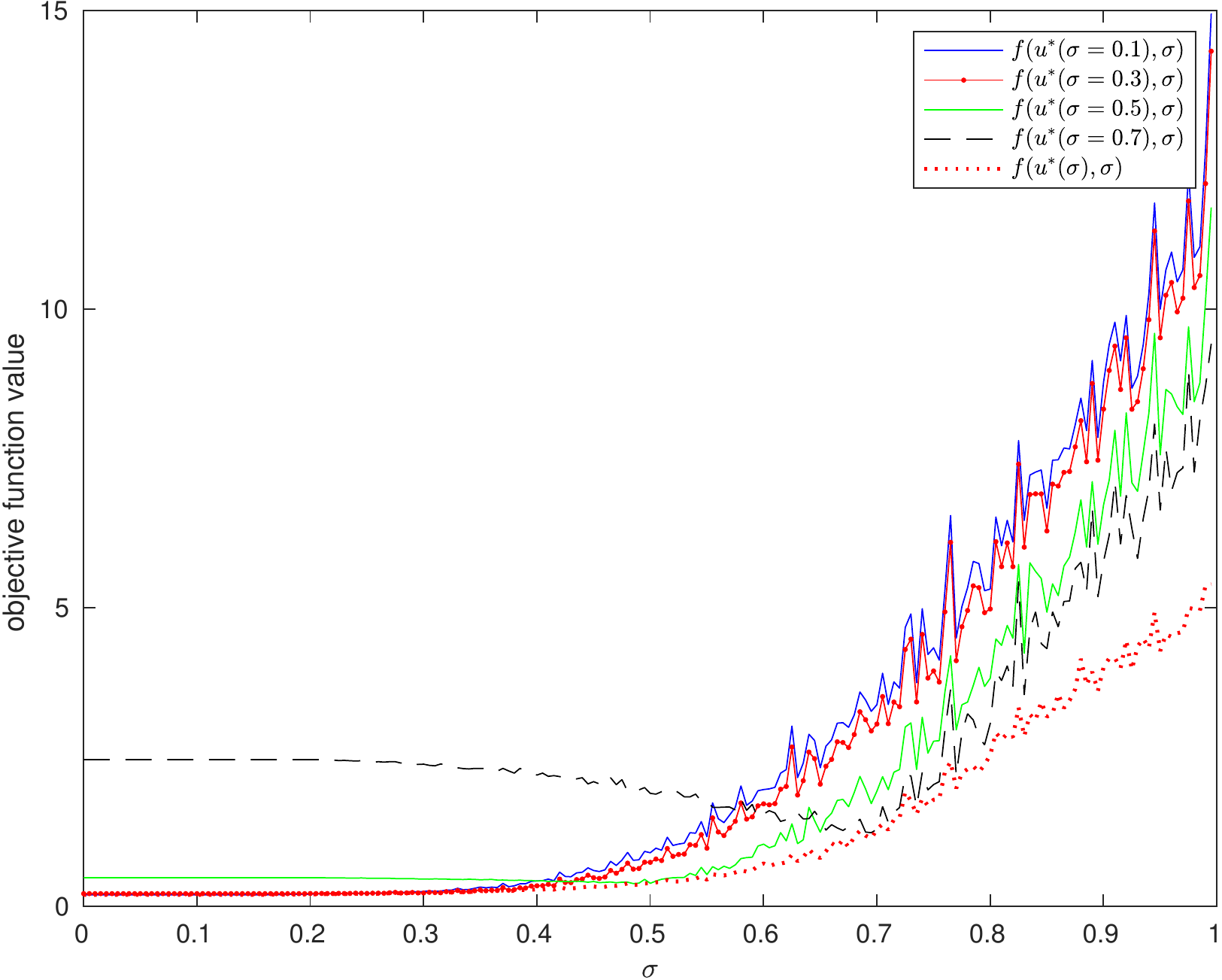}
\label{fig:WC_sens}}
\caption{Illustration of the sensitivity of the robust plans~$u^*(\sigma)$ for various~$\sigma_{\text{prior}} = \{0.1,0.3,0.5,0.7\}$.  }
\end{figure}\label{fig:sens}
In the sensitivity analysis, changes in the objective function value~$f_{\sigma}(u^*(\sigma))$ for the expected-value- and worst-case-optimization approach as a function of the standard deviation are evaluated.
In addition, it is investigated how the objective function values may change as a result of interfractional variations following a distribution different from the \textit{a priori} hypothesis. 
The results of this analysis are illustrated in Figures~4.\ref{fig:exp_sens} and~4.\ref{fig:WC_sens} for expected-value- and worst-case-optimization, respectively.
The objective function values for the worst-case-optimization approach do not appear to increase continuously with~$\sigma$. 
This appearance of noise is caused by the use of heuristics to obtain the optimal solutions~$u^*(\sigma)$ from the worst-case-optimization approach. 
This is in contrast to the values for the expected-value-optimization approach in which the optimal solutions are computed in an exact manner by exploiting its structure. 
In both optimization approaches, the objective function values for the ground truth, represented by the red dotted line, increase with the standard deviation.
This observation may indicate that the resulting treatment quality may decline as the amplitude of the variations increases. 
The results suggest that, the objective function value for the expected-value-optimization approach grows at a 'slower' rate as the standard deviation increases, while the objective function value for the worst-case-optimization approach grow at a 'faster' rate.
Concerning the sensitivity of solutions to deviations from the \textit{a priori} hypothesis, the following trends are observed.
If the distribution of the variations has been underestimated,~\textit{i.e.}~$\sigma_{\text{prior}} < \sigma$, the optimal solutions obtained from expected-value-optimization seem to result in rather small changes except for~$\sigma_{\text{prior}} = 0.1\ \text{cm}$,~\textit{i.e.} strong underestimation, where the objective function value increases the most. 
Apart from the sensitivity of solutions~$u^*(\sigma_{\text{prior}}=0.1)$ to large deviations, the increase in objective function value may be explained by higher importance weights for the CTV. 
Solutions obtained from the worst-case-optimization approach lead to very large deviations from the ground truth.
If the distribution of the variations has been overestimated,~ \textit{i.e.}~$\sigma_{\text{prior}} > \sigma$,, to a small extent, solutions from the expected-value-optimization approach result in very small deviations from the ground truth. However, the smaller the standard deviation of the actual variations, the larger the deviation in objective function value. 
These deviations are not as large as in the case of underestimation, since the importance weights for the OARs are lower than those for the CTV. 
On the other hand, solutions obtained from the worst-case-optimization approach do not appear to deviate to a great extent from the ground truth. 
An exception is given by the solution for~$\sigma_{\text{prior}} = 0.7 \text{cm}$, which represents the largest extent of overestimation in this study.
Overall, solutions of the worst-case-optimization approach appear to be more sensitive to underestimation of the distribution of interfractional variation by the \textit{a priori} hypothesis.


\subsection{Application of the Scenario-Reduction- and Tree-Generation Algorithm to Robust ART}
\begin{figure}
\center
\includegraphics[scale=0.45]{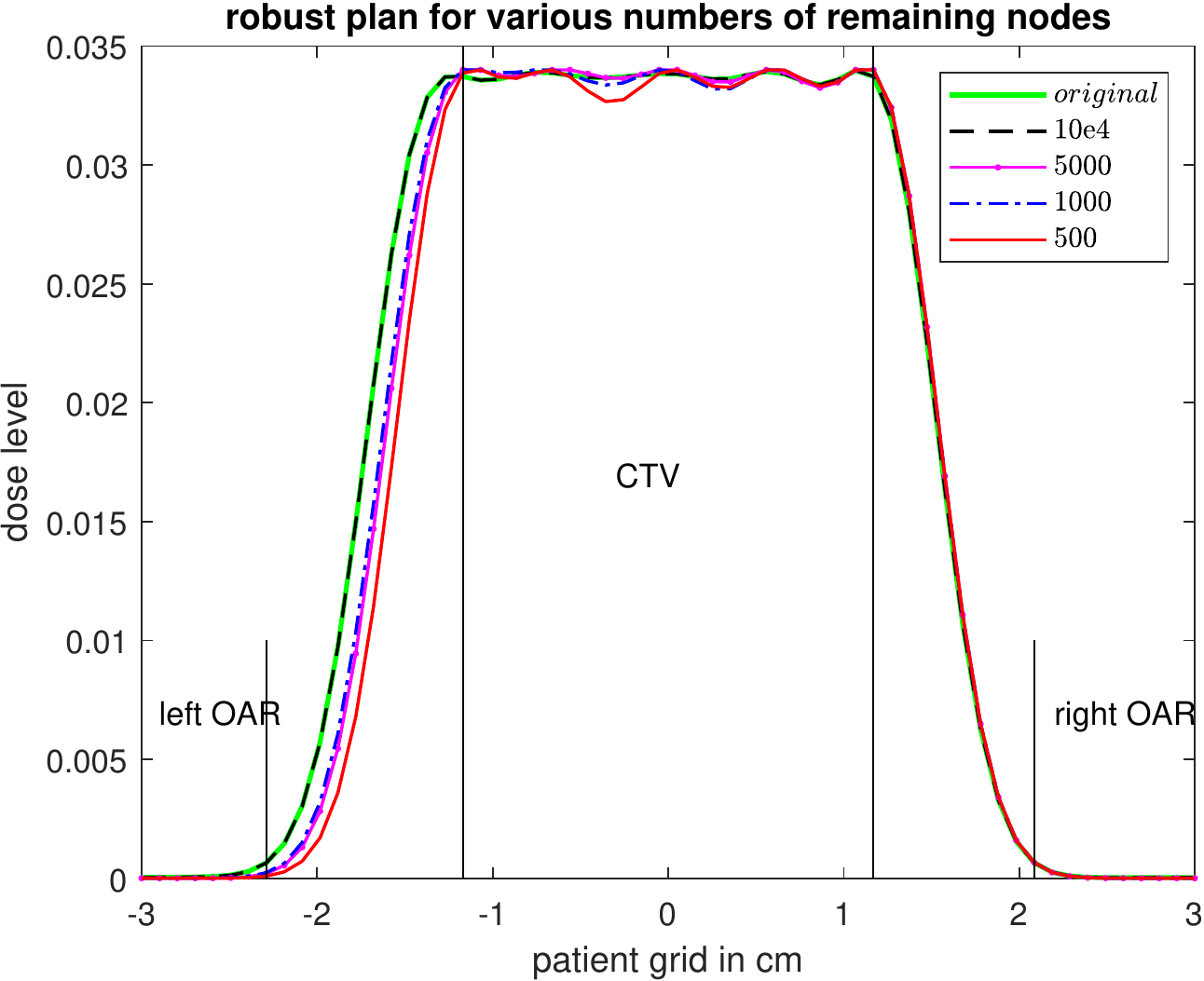}
\caption{Comparison of plan optimized with the original tree with the plans optimized for reduced trees according to~$\mathcal{S}_T = \{10.000,5000,1000,500\}$.}
\end{figure}\label{fig:redplan}
Here, the suitability of the concept of scenario-reduction for robust planning and ART is investigated.
According to the plan profiles shown in Figure~\ref{fig:redplan}, the plan of the original tree appears to be approximated best by the plan obtained from the reduced tree with~$10.000$ scenarios.
Thus, this value is used in the proposed robust adaptive framework.
However, this analysis has been carried out in an idealized setting and therefore a reduced scenario tree may be of different size for a two- or three-dimensional geometry.

\subsection{Framework for Robust Adaptive Radiation Therapy}\label{sec:Frameworkresults}

\begin{figure}
\center
\subfigure[Probability of CTV coverage is improved through adaptative replanning combined with expected-value-optimization.]{\includegraphics[scale=0.35]{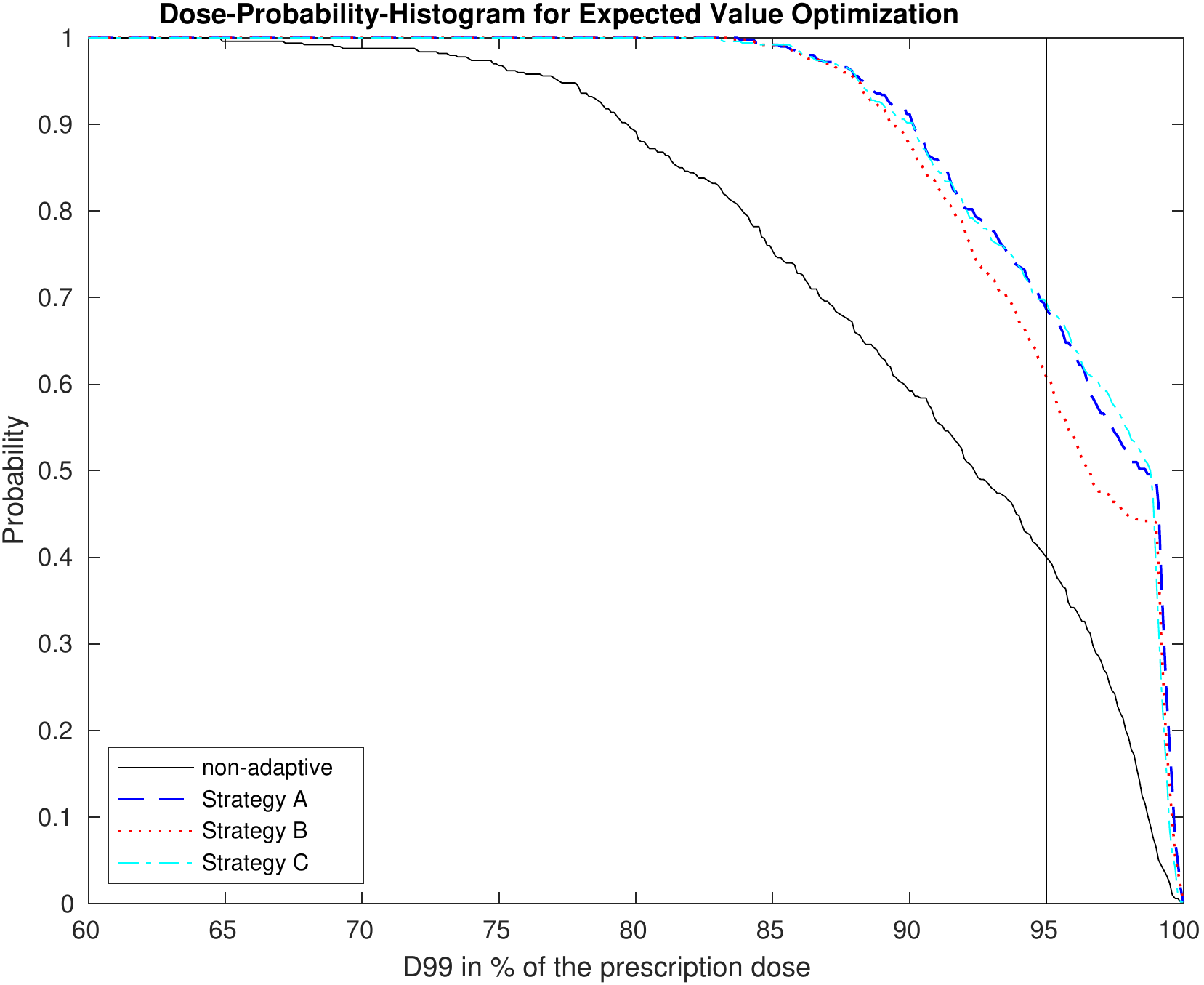}{\label{fig:CTV_exp_hist}}
}\quad
\subfigure[Probability of CTV coverage is improved through adaptative replanning combined with worst-case-optimization.]{\includegraphics[scale=0.35]{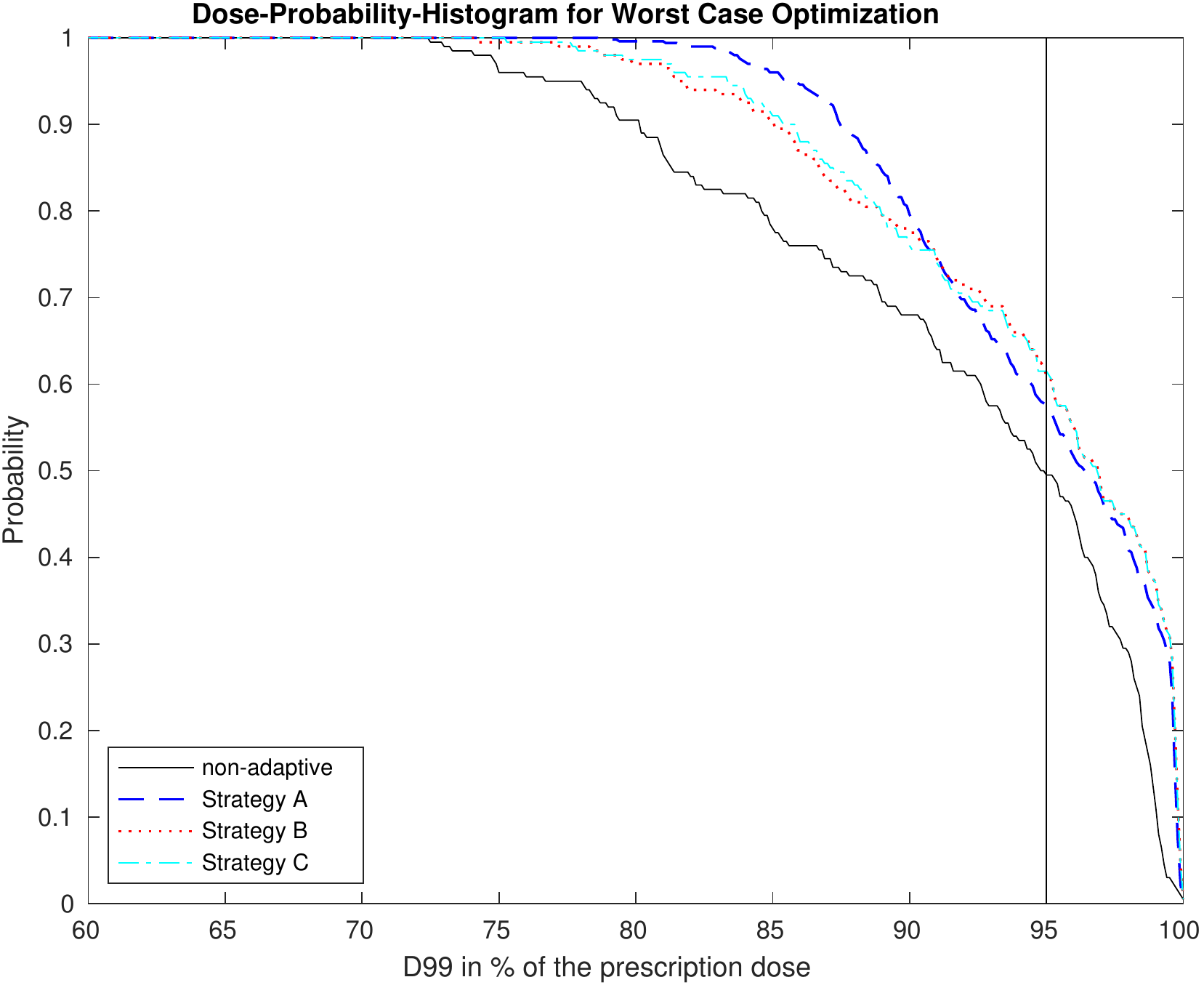}\label{fig:CTV_WC_hist}}
\caption{Comparison of the performance of the decision-support framework  using strategies A, B and C combined with expected-value- and worst-case-optimization and the corresponding non-adaptive strategies on their ability to provide certain percentage values of the prescription dose in 99\% of the CTV.}
\end{figure}

\begin{figure}
\center
\subfigure[Evaluation of the $90^{th}$ percecntile of the final accumulated dose in the right OAR as a result of non-adaptive and adaptative treatment strategies combined with expected-value-optimization.]{\includegraphics[scale=0.35]{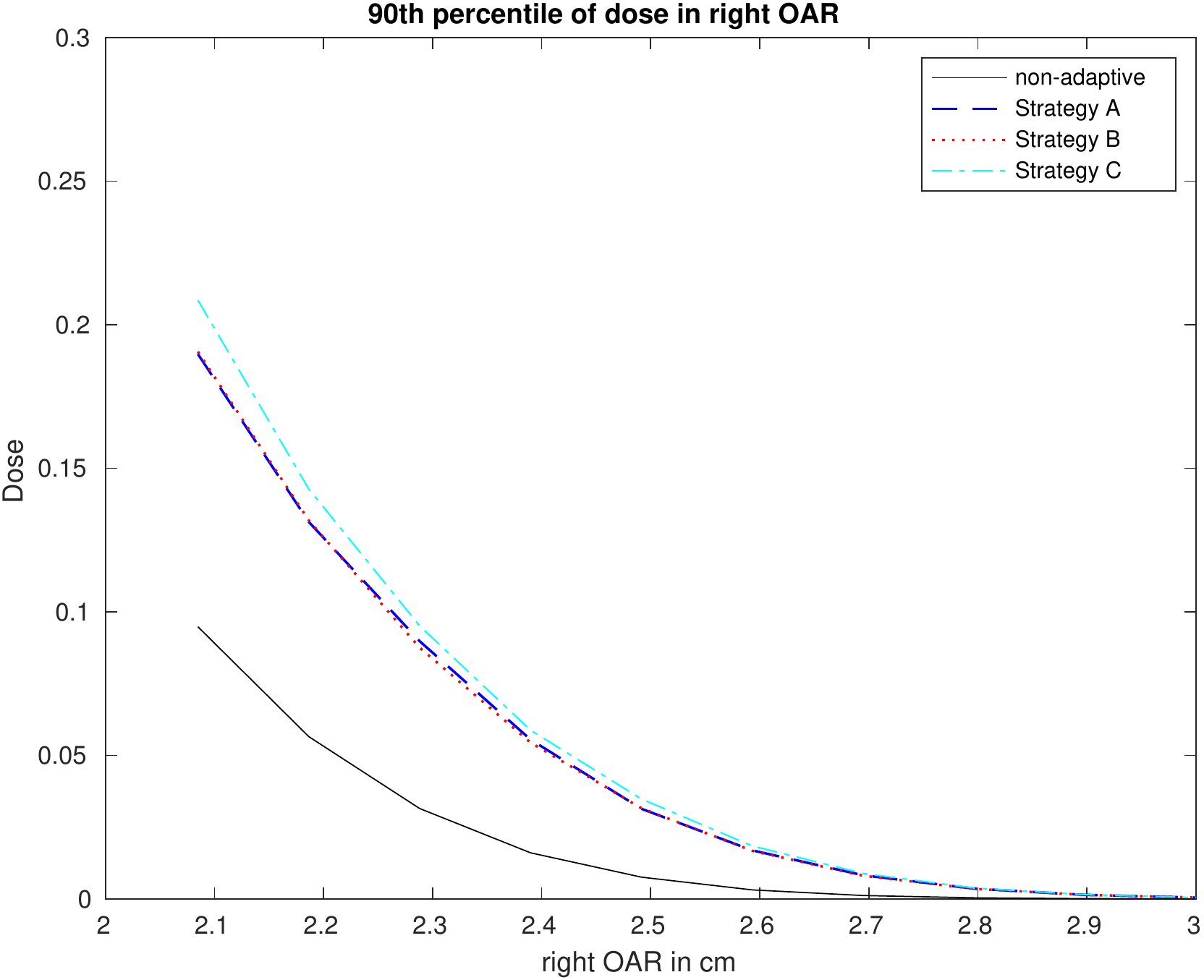}{\label{fig:OAR_exp_hist}}
}\quad
\subfigure[Evaluation of the $90^{th}$ percecntile of the final accumulated dose in the right OAR as a result of non-adaptive and adaptative treatment strategies combined with worst-case-optimization.]{\includegraphics[scale=0.35]{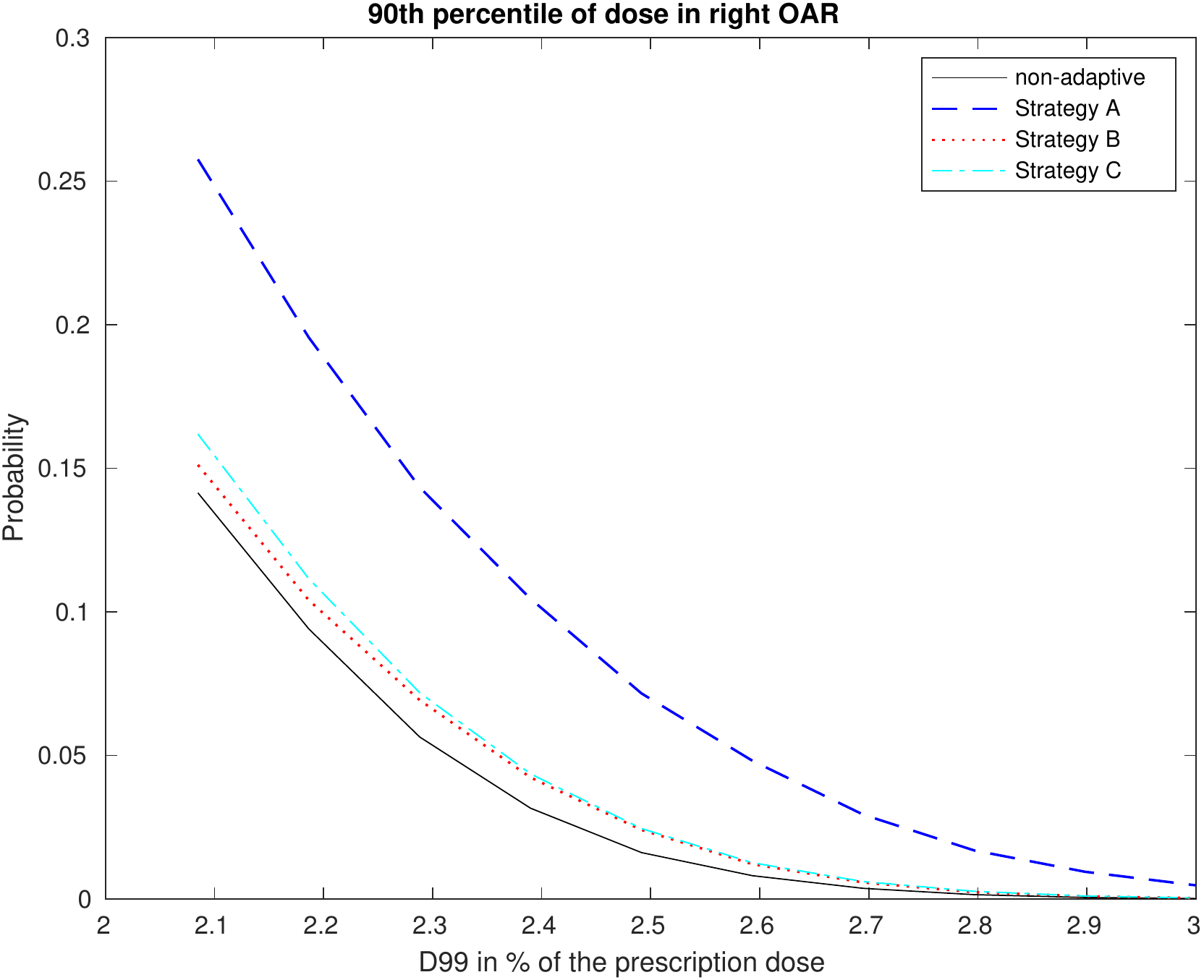}\label{fig:OAR_WC_hist}}
\caption{Comparison of the performance of the decision-support framework  using strategies A, B and C combined with expected-value- and worst-case-optimization and the corresponding non-adaptive strategies on their ability to spare dose exposure to the right OAR. We choose to show results for the right OAR, since it closer to the CTV and therefore at greater risk to receive high dose levels.}
\end{figure}

In this study, the performance of the proposed framework to handle uncertainty in simulated treatments is evaluated in  
comparison with the corresponding non-adaptive robust approach.
The performance is measured by the ability of the framework to  provide sufficient CTV coverage and protect the OARs with high probability.
CTV coverage is evaluated by using dose-probability-histograms, as shown in Figure~4.\ref{fig:CTV_exp_hist} and~4.\ref{fig:CTV_WC_hist}.
These dose-probability-histograms illustrate the probability that the final accumulated dose~$x_T$ satisfies the ICRU-guideline~\cite{ICRU62} that~99\% of the CTV will receive at least~95\% of the prescription dose. 
The probability that~99\% of the CTV will receive at least a certain dose or above is plotted along the y-axis, while the x-axis illustrates dose levels from 60\% to 100\% of the prescription dose. 
As a visual guide, the ICRU guideline is visualized by the vertical line in Figures~4.\ref{fig:CTV_exp_hist} and~4.\ref{fig:CTV_WC_hist}.
The evaluation of OAR dose exposure focuses on the right OAR, since it is closer to the CTV and therefore at greater risk of high dose exposure than the left OAR.
Dose exposure to the right OAR is evaluated by illustrating the~$90^{th}$ percentile of the final dose accumulated over simulated treatments in Figures~4.\ref{fig:OAR_exp_hist} and~4.\ref{fig:OAR_WC_hist}. 
The~$90^{th}$ percentile gives the upper threshold of dose level for 90\% of the evaluated treatments.
Thus, the overall dose exposure in the OARs for the majority of the simulated treatments can be assessed.
In addition, the suitability of the expected-value- and worst-case-optimization approach is evaluated by using Figures~\ref{fig:CTV_hist} and~\ref{fig:OAR_hist}.
Moreover, these visualized results show trends in the accumulated doses as result of employing the robust adaptive \textit{strategies~A,~B} and~\textit{C}. 
Thus, the impact of using Bayesian inference, varied adaptation frequency and adaptation triggers can be studied. 

The results in Figures~4.\ref{fig:CTV_exp_hist} and~4.\ref{fig:CTV_WC_hist} suggest that the robust adaptive framework may improve target coverage in comparison with the corresponding non-adaptive robust approach.
However, a superior adaptive adaptive strategy cannot be identified from these results.
Concerning OAR sparing, the results in Figures~4.\ref{fig:OAR_exp_hist} and~4.\ref{fig:OAR_WC_hist} suggest that \textit{strategy~B} and\textit{~C} combined with the worst-case-optimization approach lead to a relatively large decrease of dose accumulated in the right OAR.
Both \textit{strategies~B} and~\textit{C} use Bayesian inference in contrast to \textit{strategy~A}. 
Thus, this finding indicates that Bayesian inference may be a useful method for robust ART.
In contrast to \textit{strategy~B}, \textit{strategies~A} and~\textit{C} allow for adaptive replanning whenever an adaptation is triggered. 
The observation that strategy~B and~C decrease the dose to the right OAR to similar extent, suggests that an increased adaptation frequency may not be as effective as Bayesian inference in robust ART. Here, adaptation costs are lower for \textit{strategy~B} than for~\textit{C}. 
Thus, the mathematical methods used in adaptive replanning may have a greater impact on the resulting dose than increased adaptation frequency. 
Moreover, the simulations indicate that the worst-case-optimization approach may be more suitable with the Bayesian inference method.
This is in accordance with the sensitivity analysis, which suggests that the worst-case-optimization approach is more sensitive to inaccuracies in the \textit{a priori} hypothesis.
The approach to combine mathematical and user-specified adaptation triggers shows the trend for overall improvement in target coverage for both robust adaptive strategies, and of protection of the OAR for the adaptive \textit{strategies~B} and~\textit{C} combined with worst-case-optimization.
This may indicate that such an approach may be useful to have a more objective decision-making process in ART. 
Overall, these results should be read as a proof-of-concept which may be used to further develop methods for robust ART.
\label{fig:CTV_hist}
\label{fig:OAR_hist}

\section{Discussion and Outlook}\label{sec:discussion}
In this paper, a novel framework for online robust ART is introduced and evaluated.
The purpose of this novel approach is to handle interfractional geometric variations following a distribution different from the \textit{a priori} hypothesis. 
This framework is designed to address the issues of adaptation costs and computational tractability. 
Adaptation costs are addressed by three robust strategies providing Bayesian inference and/or the option of increased adaptation frequency. 
In addition, mathematical adaptation triggers are used in the proposed framework.
Computational tractability is addressed by the concept of scenario-reduction.
To the best of our knowledge, this is the first approach of its kind to online robust ART.
Therefore, this study emphasizes the mathematical properties of the proposed framework in order to understand which properties may be relevant for handling interfractional geometric variations.
For this purpose, the performance of the novel framework is evaluated in a one-dimensional phantom geometry for a series of simulated treatments.
For the sake of a fair comparison of all strategies, the same population of simulated treatments is used to analyze the performances of the robust adaptive strategies within the framework.  
During these treatments, the one-dimensional phantom geometry is subjected to rigid whole-body shifts.
In this idealized setting, computational effort is reduced as the accumulated dose can be computed and tracked in a straightforward manner. 
As a consequence of modeling interfractional geometric variations as rigid shifts, features in the dose distribution can be directly associated with changes in the corresponding optimization parameter. 
Thus, the relevant parameters and mathematical properties for an effective online robust ART framework can be identified. 
Examples for the use of a one-dimensional phantom geometry to analyze a novel approach from a mathematical perspective can be found in the literature~\cite{lof1998,chan2006,bock2017,bock2018}.
Modeling interfractional geometric vaiations as i.i.d.~shifts is reasonable for random setup errors and organ motion~\cite{vanHerk2000,unkelbach2009,unkelbach2004,fredriksson2012}.
If organ deformations were included in this study, finding the relation between features in the dose profile and the corresponding optimization parameter, may have been compromised.
Organ-deformations and an application to patient data are outside the scope of this study and postponed to a follow-up paper. 
Otherwise, characteristics of the framework in the resulting dose profiles could be overlooked or mistaken as features caused by a certain beam set-up or organ shape. 
In addition, the quadratic penalty objective function is used to exploit its convexity, smoothness and resulting globally optimal solutions.
Having globally optimal solutions in a mathematical study is useful for understanding the underlying mechanism between changing parameters and the resulting dose-profiles.
In general, the proposed framework can be combined with any type of objective function.
However, a convex objective function may be preferable in the first iteration of studying this novel framework for online robust ART. 
As a consequence of using a one-dimensional geometry, i.i.d. rigid shifts and the chosen objective function, the results of this study should be read as proof-of-concept.

Overall, the following insights are gained from this study.
First, the simulations suggest that the proposed framework may be superior in CTV coverage compared to the corresponding non-adaptive robust approach. 
Second, combining the worst-case-optimization approach with Bayesian inference appears to perform best in terms of improving CTV while decreasing dose exposure to the OAR. 
Third, the simulations indicate that adaptation costs may be controlled by employing mathematical methods such as Bayesian inference in the replanning process.
Fourth, the concept of scenario reduction has been applied to robust planning and ART and the results suggest that it may be a useful approach to address computational tractability.
This proof-of-concept study provides valuable insights into robustness, adaptation strategies and computational tractability for online robust ART.
This is crucial for further development of the proposed framework toward applying the proposed framework to clinical patient data.
\section{Conclusion}
This novel framework for online robust ART combines the strength of robust planning with the concept of ART using Bayesian inference and scenario-reduction.
The results in this study indicate that methods such as Bayesian inference may be useful to: (i) individualize plans to the actual interfractional variations of the individual case and (ii) handle adaptation costs.
In particular, the results of this proof-of-concept study suggest that the combination of the worst-case-optimization approach with Bayesian inference may lead to the largest improvements in CTV coverage and OAR protection among the studied strategies. 
Moreover, the results suggest that the concept of scenario-reduction and tree-generation may be useful for ART and robust planning in general.

\section*{Acknowledgements}
The author thanks Tatjana Pavlenko and Martina Favero for the helpful discussions on statistical methods and Bayesian inference; and Anders Forsgren and Kjell Eriksson for constructive discussions and comments throughout this study. 

\section*{References}
\bibliography{manus3bib}
\bibliographystyle{myplain}

\end{document}